\documentclass[acmsmall,screen]{acmart}
\title{InfCode-C++: Intent-Guided Semantic Retrieval and AST-Structured Search for C++ Issue Resolution}

\author{Qingao Dong}
\affiliation{%
  \institution{Beihang University}
  \city{Beijing}
  \country{China}
}
\affiliation{%
  \institution{Beijing Tokfinity Technology Co., Ltd.}
  \city{Beijing}
  \country{China}
}

\author{Mengfei Wang}
\author{Hengzhi Zhang}
\author{Zhichao Li}
\affiliation{%
  \institution{Beijing Tokfinity Technology Co., Ltd.}
  \city{Beijing}
  \country{China}
}

\author{Yuan Yuan}
\author{Mu Li}
\author{Xiang Gao}
\author{Hailong Sun}
\author{Chunming Hu}
\author{Weifeng Lv}
\affiliation{%
  \institution{Beihang University}
  \city{Beijing}
  \country{China}
}
\usepackage{xspace}
\usepackage{enumitem}
\usepackage{calc}
\usepackage{multicol} 
\usepackage{adjustbox}
\usepackage{graphicx}

\usepackage{soul}
\usepackage{tabularx}
\usepackage[noend]{algpseudocode}
\usepackage{booktabs}
\usepackage{multirow}
\usepackage{graphicx}
\usepackage{textcomp}
\usepackage{tikz}
\usepackage{xcolor}
\usepackage{enumitem}
\usepackage{listings}
\usepackage{caption}
\usepackage{alltt}
\usepackage{multirow}
\usepackage{listings}
\usepackage{color, soul}
\usepackage{textcomp}
\usepackage{framed}
\usepackage{hhline}
\usepackage{subcaption}
\usepackage[breakable]{tcolorbox}
\usepackage{float}
\usepackage{array}
\usepackage{flushend}
\usepackage{makecell}
\usepackage{balance}
\usepackage{centernot}
\usepackage[T1]{fontenc}
\usepackage[flushleft]{threeparttable}
\usepackage{url} 
\usepackage{stmaryrd}
\usepackage[colorinlistoftodos,textwidth=15mm,textsize=tiny,obeyFinal]{todonotes}
\usepackage{lstautogobble}
\usepackage{fancyvrb}
\usepackage{tcolorbox}
\usepackage{hyperref}
\usepackage{siunitx}
\usepackage{algorithm}
\usepackage{algpseudocode}

\definecolor{gray}{RGB}{215,215,215}
\definecolor{light-gray}{gray}{0.8}
\definecolor{codegreen}{rgb}{0,0.6,0}
\definecolor{codegray}{rgb}{0.5,0.5,0.5}
\definecolor{mygray}{rgb}{0.7,0.7,0.7}
\definecolor{codepurple}{rgb}{0.58,0,0.82}
\definecolor{backcolour}{rgb}{0.95,0.95,0.92}

\newcommand{\toolname}[0]{\textsc{InfCode-C++}\xspace}

\begin{abstract}
Large language model (LLM) agents have recently shown strong performance on
repository-level issue resolution, but existing systems are almost exclusively
designed for Python and rely heavily on lexical retrieval and shallow code
navigation. These approaches transfer poorly to C++ projects, where overloaded
identifiers, nested namespaces, template instantiations, and deep control-flow
structures make context retrieval and fault localization substantially more
difficult. As a result, state-of-the-art Python-oriented agents show a drastic
performance drop on the C++ subset of MultiSWE-bench.
We introduce \toolname{}, the first C++-aware autonomous system for
end-to-end issue resolution. The system combines two complementary retrieval
mechanisms---semantic code-intent retrieval and deterministic AST-structured
querying---to construct accurate, language-aware context for repair. 
These components enable precise localization and robust patch synthesis in
large, statically typed C++ repositories. Evaluated on the
\texttt{MultiSWE-bench-CPP} benchmark, \toolname{} achieves a resolution rate of
25.58\%, outperforming the strongest prior agent by 10.85 percentage points and
more than doubling the performance of MSWE-agent. Ablation and behavioral
studies further demonstrate the critical role of semantic retrieval,
structural analysis, and accurate reproduction in C++ issue resolution.
\toolname{} highlights the need for language-aware reasoning in multi-language
software agents and establishes a foundation for future research on scalable,
LLM-driven repair for complex, statically typed ecosystems.
\end{abstract}

\begin{document}
\maketitle

\section{INTRODUCTION}
\label{sec:intro}

Automated software issue resolution stands as a critical challenge in software engineering, promising to significantly reduce developer workload and accelerate maintenance cycles. The emergence of comprehensive benchmarks, most notably SWE-bench \cite{swebench}, which evaluates agents on thousands of real-world GitHub issues, has been pivotal in measuring progress. On this benchmark, which is predominantly composed of Python repositories, state-of-the-art LLM-powered agents like Trae Agent have demonstrated impressive capabilities, achieving resolution rates as high as 75.20\% \cite{trae}. However, this success in memory-safe, dynamically-typed languages masks a significant performance gap in other critical domains. Specifically, resolving issues in large-scale, high-performance systems written in C++ remains a formidable, unsolved challenge. Recent evaluations on multilingual benchmarks, such as MultiSWE-bench \cite{multiswebench}, highlight this disparity; a leading agent configuration, Mopenhands paired with Claude 3.7 Sonnet, achieved a resolution rate of only 14.7\% on C++ tasks \cite{openhands}. This stark performance drop underscores that current methodologies are ill-equipped to handle the distinct complexities of C++, necessitating a shift from general-purpose approaches to language-specific, structurally-aware solutions.

\textbf{Ambiguous Context Localization without Precise Artifacts.}
A primary challenge in C++ issue resolution is ambiguous context localization, particularly when the issue description lacks precise artifacts such as a stack trace. This problem is exacerbated for non-crashing bugs, such as logical errors or performance bottlenecks, where no immediate fault signal is available. Unlike a crash bug which pinpoints a faulting location, these issues require the agent to deduce the relevant code context from a natural language description alone. In a sprawling, million-line C++ codebase, this task becomes a ``needle in a haystack'' problem. Existing approaches~\cite{agentless,sweagent,openhands} often resort to rudimentary heuristics, attempting to guess relevant file, class, or function names based on superficial lexical cues in the issue report. This approach is inefficient, unreliable, and frequently retrieves irrelevant context, forcing the LLM to analyze hundreds or even thousands of lines of unrelated code, thus leading to incorrect or incomplete patches~\cite{chen2025revisit}.

\textbf{Insufficiency of Lexical Search in Complex Code Structures.}
This localization difficulty is compounded by the inadequacy of standard lexical search tools. Many contemporary agents rely on `grep' utilities for context retrieval~\cite{trae,bugpilot,sweagent,openhands}. While effective for simple pattern matching, these tools are fundamentally limited and ill-suited for C++. The structural complexity of C++, characterized by features such as namespaces, function and operator overloading, intricate template metaprogramming, and deep inheritance hierarchies—renders text-based search insufficient. For example, a `grep' query for a function named `update' cannot semantically disambiguate between `UI::update()', `Database::update()', and a similarly named method in a base class. It also fails to resolve virtual function calls or understand template instantiations. This lexical ambiguity pollutes the context provided to the LLM with a high volume of irrelevant and misleading information, obscuring the true semantic relationships and call graphs, and ultimately hindering its ability to perform robust bug resolution.

To address these specific challenges in C++ repositories, we propose a novel, two-pronged approach that enhances an LLM agent's ability to understand and navigate complex C++ codebases. Our method combines a high-level, semantic retrieval strategy based on code intent with a low-level, precise structural search mechanism based on the Abstract Syntax Tree (AST).

First, we introduce a semantic context retrieval framework based on \textit{code intent}. This method moves beyond simple file-based retrieval by associating disparate software artifacts, such as files, classes, and functions, that collectively implement a specific, high-level feature. This association allows an LLM agent to first query the codebase using natural language to identify relevant functional modules (e.g., ``locate components responsible for data serialization''). Once this high-level context is established, the agent can execute more granular queries to inspect the specific code blocks necessary for the fix, effectively narrowing the search space from the entire repository to a few relevant components.

Second, to ``go beyond grep'' and overcome the limitations of lexical search, we develop an AST-based structured query engine. This engine parses the C++ codebase into its abstract syntax tree, enabling precise queries based on code structure rather than text. We expose this capability to the LLM agent through a suite of robust tools, including `FindClass', `FindFunction', and `GetInheritanceChain'. These tools empower the agent to navigate C++'s complex structures deterministically, find precise definitions, and understand critical relationships (e.g., class hierarchies or virtual function overrides) that are invisible to text-based tools.

We thus present \toolname{}, an autonomous agent that integrates these two capabilities to systematically resolve C++ issues. Starting from a natural language issue description, \toolname{} first employs the code-intent semantic retrieval to perform a high-level semantic search, identifying a small set of candidate functional modules relevant to the issue. Subsequently, the agent deploys its AST-based search tools to conduct a precise, structural analysis within these modules. This two-stage process—broad semantic filtering followed by deep structural validation—allows the agent to resolve ambiguities like function overloading and namespace conflicts. This procedure provides the LLM with a concise, accurate, and semantically-rich context that is not merely lexically relevant but structurally correct, facilitating the construction of a robust patch.

The main contributions of this work are as follows:
\begin{itemize}
    \item \textbf{Semantic Code-Intent RAG:} We propose a novel semantic retrieval framework that maps high-level functional descriptions (the intent) to relevant, multi-file code artifacts. This allows an agent to bypass ambiguous lexical search and instead retrieve context based on feature-level understanding.
    
    \item \textbf{Structural-Aware AST Querying:} We develop a set of agent-usable tools that operate directly on the C++ AST. This mechanism, featuring tools like `FindClass' and `GetInheritanceChain', overcomes the limitations of text-based search by providing deterministic, semantically-correct navigation of complex C++ structures.
    
    \item \textbf{High-Efficacy Agent and Public Artifact:} We design and implement \toolname{}, an agent integrating our two-pronged retrieval system. Our tool shows high efficacy, solving \textbf{25.58\%} of issues on the C++ subset of MultiSWE-bench \cite{multiswebench}, significantly outperforming prior state-of-the-art. To support the research community and facilitate future work, we release \toolname{} and our evaluation benchmark as an open-source project at \url{https://github.com/Tokfinity/InfCode}.
\end{itemize}

\section{Motivation Example}
\label{sec:motivation}

\begin{figure}[t]
    \centering
    \includegraphics[width=0.98\linewidth,
                     trim=0 0 40pt 0, clip]{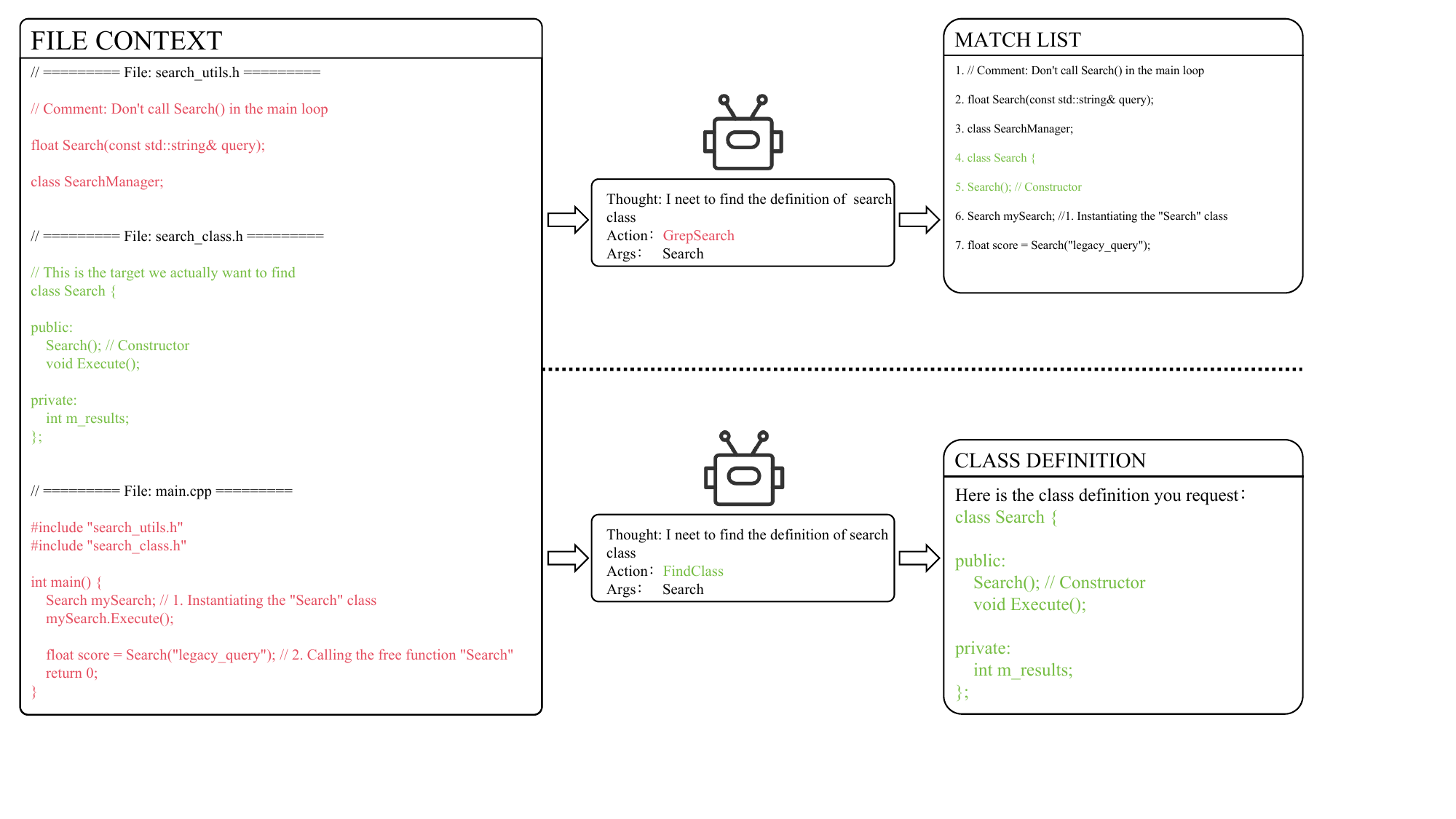}
    \caption{Motivation example illustrating the ambiguity of lexical retrieval
    and the correctness of AST-structured querying in C++ codebases.}
    \label{fig:motivation_example}
\end{figure}

We present a motivating example that highlights the core challenges of C++ issue
resolution and the necessity of the retrieval mechanisms introduced in
\toolname{}. The example is adapted from a real-world multi-file C++ project,
and its structure is illustrated in Figure~\ref{fig:motivation_example}.

\paragraph{Lexical retrieval fails in the presence of overloaded identifiers.}
The file context shown on the left side of the figure contains three files:
\texttt{search\_utils.h}, which defines a free function \texttt{Search}; 
\texttt{search\_class.h}, which defines a class \texttt{Search} with a constructor
and a member function; and \texttt{main.cpp}, which uses both the class
\texttt{Search} and the free function \texttt{Search}.  
This mixture of identically named entities is common in C++ due to overloaded
functions, namespace scoping, forward declarations, and project-specific naming
conventions.

When an agent attempts to locate the definition of ``the \texttt{Search} class,''
a lexical retrieval tool such as \texttt{grep} matches every appearance of the
string ``Search,'' producing the heterogeneous match list in the top-right of
Figure~\ref{fig:motivation_example}.  
The list includes comments, forward declarations, free functions, class names,
constructor invocations, and call sites.  
Crucially, lexical search cannot distinguish between:

\begin{itemize}[leftmargin=*]
    \item a free function \texttt{Search(const std::string\&)},
    \item a forward declaration \texttt{class SearchManager},
    \item the actual class definition \texttt{class Search \{...\}}, and
    \item uses of the class constructor or variable instantiations.
\end{itemize}

As a result, lexical tools provide an over-approximate and noise-dominated
context. The agent must manually filter a large number of irrelevant matches
to locate the true class definition.

\paragraph{Structured retrieval resolves ambiguity.}
With our structural querying mechanism, the agent instead issues:

\[
\texttt{FindClass("Search")}
\]

which triggers a deterministic AST-level lookup over the parsed project.  
As shown in the bottom-right of Figure~\ref{fig:motivation_example}, the
AST-based query resolves all identifier collisions and directly returns the
precise class definition, correctly handling overloaded names, constructor
declarations, and unrelated function symbols.  

This example demonstrates why C++ presents significantly greater retrieval
ambiguity than languages such as Python.  
Identifier overloading, namespace scoping, and multiple declaration forms make
lexical search fundamentally insufficient for fault localization.  
By combining intent-based semantic filtering and AST-structured queries,
\toolname{} eliminates lexical ambiguity and provides the agent with an
accurate, noise-free program representation—an essential capability for
resolving complex C++ issues.

\section{METHODOLOGY}
\label{sec:method}

\subsection{Problem Definition}

Given a software repository $C$ and a natural-language issue description $D$, the objective is to synthesize a patch $p$ such that the updated repository $C' = C \oplus p$ satisfies the behavioral requirements expressed in $D$ while preserving the original functionality. Let $T$ denote the regression test suite that accompanies the repository. A correct patch must satisfy:

\[
t(C') = t(C), \; \forall t \in T
\quad\text{and}\quad
\exists\, t_D: t_D(C') = \texttt{PASS},
\]

where $t_D$ is a failing test case that captures the erroneous behavior described in $D$.

This task is especially challenging in C++ repositories due to structural complexity, overloaded identifiers, namespace shadowing, templates, and deep inheritance hierarchies, which significantly hinder reliable context retrieval and localization. \toolname{} addresses these challenges through a structured, multi-agent design.

\subsection{System Overview}

\toolname{} is implemented as a modular, autonomous multi-agent framework comprising three cooperating agents:  
(1) the \textit{Reproducer Agent},  
(2) the \textit{Patch Agent}, and  
(3) the \textit{Selector Agent}.  

\begin{figure*}[t]
    \centering
    \includegraphics[width=\linewidth, trim=20pt 40pt 20pt 40pt, clip]{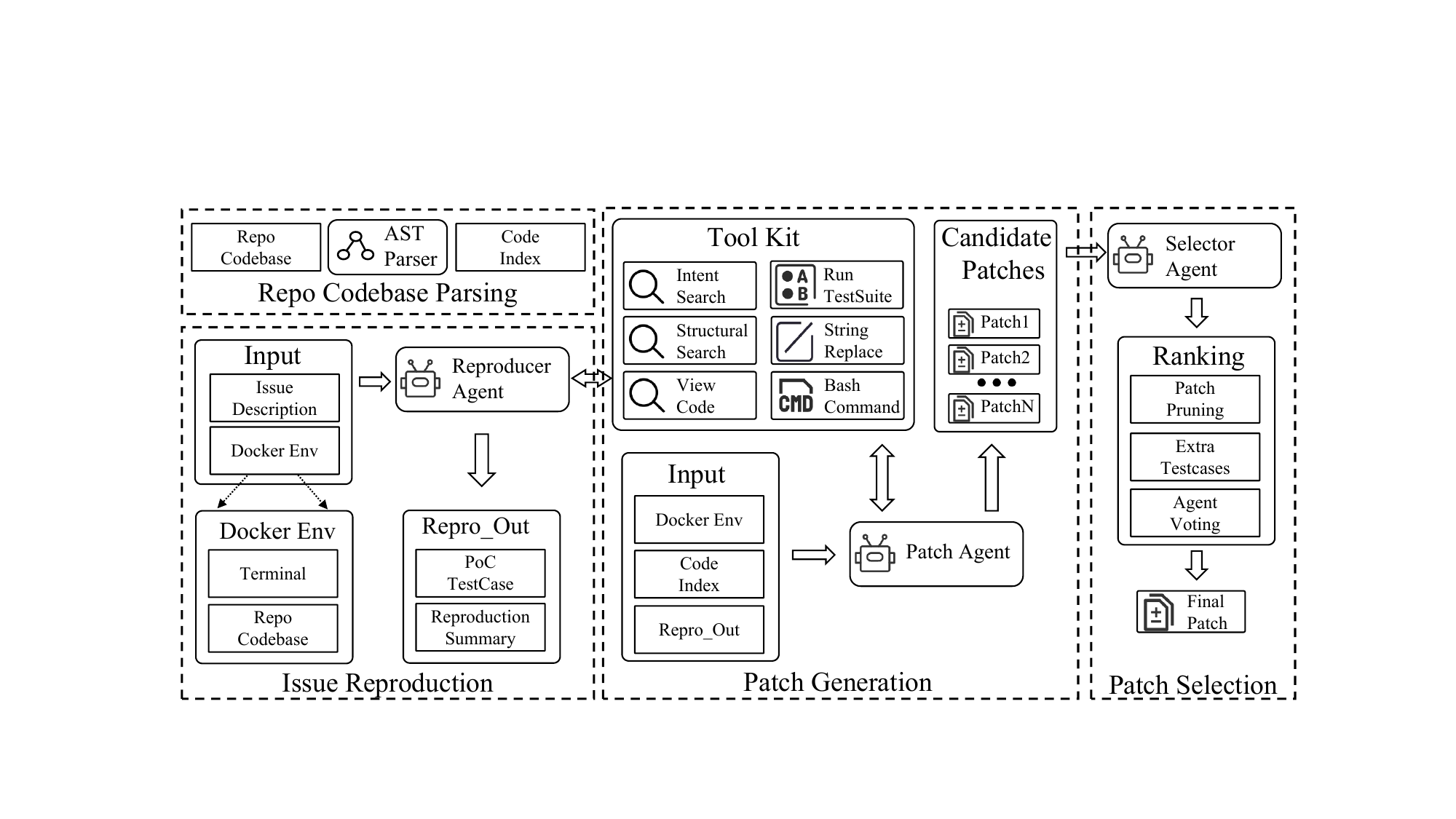}
    \caption{Overall system architecture of \toolname{}. 
    The framework consists of three collaborative agents: 
    (1) the Reproducer Agent, which reconstructs the failure and produces a concrete reproducing test; 
    (2) the Patch Agent, which performs semantic intent retrieval, AST-structured querying, 
    and multi-step patch synthesis using a rich tool kit; 
    (3) the Selector Agent, which prunes, validates, and ranks candidate patches through 
    regression testing, extra test generation, and agent voting.}
    \label{fig:architecture}
\end{figure*}

The system architecture includes four main stages—\textit{Repository Parsing}, \textit{Issue Reproduction}, \textit{Patch Generation}, and \textit{Patch Selection}—which correspond directly to the workflow illustrated in Figure~\ref{fig:architecture}.

\begin{itemize}[leftmargin=*]
    \item \textbf{Repository Parsing:} Prepares the repository by constructing an AST-based structural index and a semantic code-intent index to support subsequent retrieval.
    \item \textbf{Issue Reproduction:} Converts the natural-language issue description into an executable proof-of-concept (PoC) test case and a reproduction summary.
    \item \textbf{Patch Generation:} Performs C++-aware context retrieval—combining semantic intent search with AST structural queries—to localize faults and synthesize candidate patches.
    \item \textbf{Patch Selection:} Performs a hierarchical patch selection process that begins with 
    patch pruning, which merges semantically equivalent candidates and removes formatting-only or 
    comment-only edits; proceeds to behavioral validation, where the Selector Agent generates 
    additional test cases to filter out fault-preserving patches; and concludes with an 
    agent-voting mechanism that ranks the remaining candidates and selects the final patch.
    
\end{itemize}

This pipeline allows \toolname{} to combine high-level semantic reasoning with low-level structural precision, enabling robust fault localization and patch synthesis in complex C++ repositories.

\subsection{Repository Parsing}

Before issue resolution begins, the repository is statically analyzed to build two retrieval structures essential for C++ code navigation:

\paragraph{AST-Based Structural Index.}
The repository is parsed into an abstract syntax tree:
\[
\mathcal{T}_C = (\mathcal{N}, \mathcal{E}),
\]
where nodes $\mathcal{N}$ represent syntactic constructs (e.g., classes, methods, template instantiations) and edges $\mathcal{E}$ capture structural relations such as inheritance, call edges, containment, and overload groups. This index enables deterministic structural queries that avoid ambiguities inherent to lexical search.

\paragraph{Semantic Code-Intent Index.}
We further construct a semantic embedding index that maps feature-level intents to code artifacts:
\[
\mathcal{I}_{\text{intent}} : \texttt{Intent} \mapsto \{A_1, A_2, \ldots, A_k\},
\]
where each $A_i$ is a file, class, or function participating in the implementation of a high-level feature. This index enables intent-driven retrieval of relevant subsystems.

Together, these two indices form the core of \toolname{}’s C++-aware retrieval toolkit.

\subsection{Issue Reproduction}

The \textit{Reproducer Agent} operationalizes the issue description by producing an executable test case $t_D$ that deterministically reproduces the failure. This requires extracting execution conditions, inputs, expected and observed behaviors, and environmental dependencies.

The agent interacts with a sandboxed Docker environment containing the repository snapshot and produces:

\begin{itemize}[leftmargin=*]
    \item \textbf{PoC Test Case} capturing the erroneous behavior.
    \item \textbf{Reproduction Summary} including execution traces and environment details.
\end{itemize}

This stage provides a concrete behavioral oracle for validating and ranking candidate patches in later stages.

Formally, the reproducer defines a partial behavioral specification:
\[
\texttt{Exec}(C, t_D) = \texttt{Fail}.
\]

\subsection{Patch Agent: Context Retrieval and Patch Generation}

The \textit{Patch Agent} is responsible for fault localization and candidate patch generation. Unlike Python-focused agents, which often rely on lexical retrieval, \toolname{} integrates two complementary retrieval mechanisms specifically designed for C++.

\subsubsection{Semantic Retrieval via Code Intent}

Given the issue description $D$, the agent constructs a query $q$ and invokes:
\[
\texttt{QueryCodeIntent}(q) \rightarrow C_{\text{intent}},
\]
where $C_{\text{intent}}$ is a semantically coherent subset of the repository. This step identifies the subsystem implementing the relevant feature(s). Since:

\[
|C_{\text{intent}}| \ll |C|,
\]

this significantly eliminates irrelevant modules before structural reasoning.

\subsubsection{Structural Retrieval via AST Querying}

Within $C_{\text{intent}}$, the agent performs precise structural analysis by issuing queries such as:

\begin{itemize}[leftmargin=*]
    \item \texttt{FindClass(className)} \quad (locates class definitions across nested namespaces)
    \item \texttt{FindFunction(spec, name)} \quad (locates function definitions within the specified scope and resolves overloads)
    \item \texttt{GetInheritanceChain(className)} \quad (extracts base and derived class hierarchies)
    \item \texttt{GetFunctionCalls(className, functionName)} \quad (computes intra-class and inter-module call graphs)
\end{itemize}

Because these operations rely on AST structure rather than surface tokens, they handle overloaded identifiers, template instantiation, and nested namespaces—scenarios where lexical baselines fail.

The intersection of semantic and structural evidence yields the localized fault region:
\[
L = \mathcal{I}_{\text{intent}}(D) \cap \mathcal{G}_{\text{bug}},
\]
where $\mathcal{G}_{\text{bug}}$ is the structurally derived defect subgraph.

The Patch Agent then synthesizes candidate patches $\{p_1,\ldots,p_n\}$ conditioned on $(D, L, C_L)$.

\subsection{Patch Selection}

The \textit{Selector Agent} performs a hierarchical refinement procedure to identify the final patch 
from the set of candidates $\{p_1, \ldots, p_n\}$ generated by the Patch Agent. 
This procedure consists of three stages: patch pruning, behavioral validation, and agent voting.

\paragraph{Patch Pruning.}
Given a candidate patch $p_i$, the Selector first applies a pruning function
$\texttt{Prune}(\cdot)$ that removes edits that do not affect program behavior.
This includes formatting-only changes, comment-only modifications, and 
semantically equivalent patches that differ only in syntactic form.
The pruning stage reduces redundancy and eliminates candidates that cannot 
contribute to behavioral correction. Formally, pruning yields:
\[
\widehat{p}_i = \texttt{Prune}(p_i),
\]
and only behaviorally distinct patches are retained for further evaluation.

\paragraph{Behavioral Validation.}
For each pruned candidate $\widehat{p}_i$, the Selector applies the patch to the repository:
\[
C'_i = C \oplus \widehat{p}_i.
\]
The agent then validates $C'_i$ using both the reproducer test $t_D$ and additional
Selector-generated test cases $\mathcal{T}_{\mathrm{extra}}$:
\[
t_D(C'_i) = \texttt{PASS}
\quad\wedge\quad
\forall t \in \mathcal{T}_{\mathrm{extra}},\; t(C'_i) = \texttt{PASS}.
\]
Candidates failing any test are discarded. The surviving set is denoted
$\mathcal{P}_{\text{valid}}$.

\paragraph{Agent Voting.}
If multiple candidates remain, the Selector triggers an LLM-based voting 
mechanism that scores each patch according to its semantic alignment with the issue description $D$,
its minimality, and its consistency with the localized defect region.
The final patch is selected as:
\[
p_{\text{final}} = 
\arg\max_{p_i \in \mathcal{P}_{\text{valid}}}
\texttt{VoteScore}(p_i, D).
\]

This hierarchical design ensures that non-impactful edits are removed early, 
fault-preserving patches are filtered through behavioral validation, 
and the final decision reflects semantically grounded reasoning rather than syntactic similarity alone.

\subsection{End-to-End Execution Flow}

The complete workflow of \toolname{} follows a deterministic pipeline:

\[
(C, D)
\;\xrightarrow{\text{Reproducer Agent}}\; t_D
\;\xrightarrow{\text{Patch Agent}}\; \{p_1, \ldots, p_n\}
\;\xrightarrow{\text{Selector Agent}}\; p_{\text{final}}.
\]

By combining semantic intent analysis, AST-based structural reasoning, and 
hierarchical patch refinement, \toolname{} achieves robust issue resolution for structurally complex C++ repositories.

\begin{algorithm}[t]
\caption{End-to-End Workflow of \toolname{}}
\label{alg:workflow}
\begin{algorithmic}[1]
\Require 
    Software codebase $C$, 
    Issue description $D$,
    Regression test suite $T$
\Ensure 
    Validated patch $p_{\text{final}}$ satisfying $D$ and preserving $T$

\Statex
\Function{Reproduce}{$C, D$}
    \State Parse $D$ to extract input configuration, triggering behavior, and expected outcome
    \State Generate executable reproducer test $t_D$
    \State Verify $\texttt{Exec}(C, t_D) = \texttt{Fail}$
    \State \Return $t_D$
\EndFunction

\Statex
\Function{Patch}{$C, D, t_D$}
    \State $C_{\text{intent}} \gets$ \Call{QueryCodeIntent}{$D$}  \Comment{Semantic retrieval}
    \State $\mathcal{T}_C \gets$ \Call{ParseToAST}{$C_{\text{intent}}$} \Comment{AST construction}
    \State $L \gets$ \Call{LocateBug}{$\mathcal{T}_C, D$}          \Comment{Structural localization}
    \State $\{p_1, \ldots, p_n\} \gets$ \Call{GeneratePatches}{$L, D$}
    \State \Return $\{p_1, \ldots, p_n\}$
\EndFunction

\Statex
\Function{Select}{$C, T, t_D, \{p_1, \ldots, p_n\}$}
    \State $\mathcal{P}_{\text{valid}} \gets \emptyset$
    \For{$i \gets 1$ \textbf{to} $n$}
        \State $p_i \gets p_i$  \Comment{Candidate patch}
        \State $C'_i \gets C \oplus p_i$
        \If{$t_D(C'_i) = \texttt{PASS}$ \textbf{and} $\forall t \in T:\, t(C'_i) = t(C)$}
            \State $\mathcal{P}_{\text{valid}} \gets \mathcal{P}_{\text{valid}} \cup \{p_i\}$
        \EndIf
    \EndFor
    \If{$\mathcal{P}_{\text{valid}} = \emptyset$}
        \State \Return \texttt{FAILURE}
    \Else
        \State $p_{\text{final}} \gets \arg\min_{p_i \in \mathcal{P}_{\text{valid}}} \texttt{Complexity}(p_i)$
        \State \Return $p_{\text{final}}$
    \EndIf
\EndFunction

\Statex
\State $t_D \gets$ \Call{Reproduce}{$C, D$}
\State $\{p_1, \ldots, p_n\} \gets$ \Call{Patch}{$C, D, t_D$}
\State $p_{\text{final}} \gets$ \Call{Select}{$C, T, t_D, \{p_1, \ldots, p_n\}$}
\State \Return $p_{\text{final}}$
\end{algorithmic}
\end{algorithm}

The complete procedure of \toolname{} is summarized in Algorithm~\ref{alg:workflow}.

\section{EXPERIMENTAL SETUP}
\label{sec:experiment}

\subsection{Benchmark}

We evaluate \toolname{} on the C++ subset of the MultiSWE-bench benchmark, denoted as \texttt{MultiSWE\-bench\-CPP}. 
As shown in Table~\ref{tab:cpp_repo_stats}, the benchmark contains 129 real-world GitHub issues from actively maintained C++ repositories. 
Each instance provides a natural-language issue description, a code snapshot, and a regression test suite. 
Agents must synthesize a patch that satisfies the behavioral requirements described in the issue report while preserving all existing behaviors validated by the test suite.

All 129 instances are drawn from five actively maintained C++ libraries, whose statistics are
summarized in Table~\ref{tab:cpp_repo_stats}. The repositories range from \texttt{fmtlib/fmt} with
25 files and 36.4k lines of code to \texttt{nlohmann/json} and \texttt{simdjson/simdjson} with
477 and 455 files and 124.7k and 229.7k lines of code respectively, while \texttt{catchorg/Catch2}
and \texttt{yhirose/cpp-httplib} contribute additional large, header-heavy codebases. This scale
and modularity substantially increase the search space for localization compared to typical Python
benchmarks. All instances include ground-truth patches, enabling objective and reproducible
correctness evaluation. While some issue descriptions provide stack traces or partial execution
logs, such information is inconsistent and often incomplete in these C++ projects (due to inlining,
template instantiation, and namespace resolution), so agents cannot rely on auxiliary execution
signals and must localize faults primarily from the natural-language description and the codebase
itself.

\begin{table}[t]
\centering
\small
\caption{Repository statistics of the \texttt{Multi\-SWE\-bench\-CPP} benchmark.}
\label{tab:cpp_repo_stats}
\begin{tabular}{lccc}
\toprule
\textbf{Repository} & \textbf{Files} & \textbf{Locs} & \textbf{Instances} \\
\midrule
catchorg/Catch2         & 399 & 58.0k  & 12 \\
fmtlib/fmt              & 25  & 36.4k  & 41 \\
nlohmann/json           & 477 & 124.7k & 55 \\
simdjson/simdjson       & 455 & 229.7k & 20 \\
yhirose/cpp-httplib     & 33  & 50.9k  & 1 \\
\bottomrule
\end{tabular}
\end{table}

\subsection{Baselines}

We compare \toolname{} with three state-of-the-art agent-based repair systems:
Mopenhands~\cite{openhands}, MSWE-Agent~\cite{sweagent}, and MAgentless~\cite{agentless}. 
These systems represent the strongest publicly available baselines on MultiSWE-bench, covering tool-driven, hybrid, and model-centric repair strategies.

\subsection{Evaluation Protocol}

All methods are evaluated using the official MultiSWE-bench pipeline. 
A patch is considered correct if and only if it passes both the issue-specific reproducer test and the full regression test suite. 
Each issue is evaluated independently, and each agent is given a single attempt per instance to ensure strict comparability.

\subsection{Agent Configuration}

\toolname{} is evaluated under two backend large language models: 
GPT-5-20250807~\cite{openai_gpt5_2025} and DeepSeek-V3~\cite{liu2024deepseek}. 
The system adopts a three-agent configuration consisting of the Reproducer Agent, the Patch Agent, and the Selector Agent.

\paragraph{Reproducer Agent.}
This agent converts the natural-language issue description into an executable reproducer script.
We set a maximum interaction budget of 20 LLM turns to balance generation complexity and stability.  

\paragraph{Patch Agent.}
The Patch Agent performs context retrieval and patch synthesis using the full pipeline in Section~\ref{sec:method}, which integrates semantic code-intent retrieval and AST-structured searching.  
We allow up to 50 LLM turns for this agent, enabling multi-step reasoning across semantic and structural contexts.  
During test-time scaling, the agent generates 10 candidate patches conditioned on the localized context.  
These candidates are subsequently filtered, pruned, and validated in later stages of the pipeline.

\paragraph{Selector Agent.}
The Selector Agent receives all candidate patches from the Patch Agent and conducts behavioral validation.  
It executes candidate patches against the reproducer test and regression suite, prunes semantically equivalent or comment-only edits, and applies agent voting to identify the final patch.  
The Selector Agent does not require additional LLM turns beyond those used for ranking and semantic comparison during patch selection.

\paragraph{Baselines.}
We directly adopt the published results for all competing systems without modifying their agent configurations, toolchains, prompt formats, or execution constraints.  
This ensures that our comparisons faithfully reflect the behavior of each system under the official MultiSWE-bench evaluation protocol.

\subsection{Metric}

We report \textit{issue resolution rate}, defined as the proportion of issues for which an agent produces a correct patch under the evaluation protocol outlined above. 
This is the primary metric used by MultiSWE-bench and reflects the end-to-end repair capability of each system.

\subsection{Research Questions}

We structure our empirical study around the following three research questions.

\textbf{RQ1: How effectively does \toolname{} resolve C++ issues on the \texttt{MultiSWE-bench-CPP} benchmark?}

This research question evaluates the end-to-end repair capability of \toolname{} under the official MultiSWE-bench evaluation protocol. 
It measures whether the proposed retrieval and reasoning mechanisms improve issue resolution rates relative to existing state-of-the-art agents. 
This RQ establishes the overall effectiveness of our system in real-world C++ repair settings.

\textbf{RQ2: What is the contribution of each component in \toolname{}'s retrieval and reasoning pipeline?}

This research question examines the impact of the system's key components through ablation experiments. 
We isolate the effects of semantic code-intent retrieval and AST-structured querying, and assess the performance of the Reproducer Agent and Patch Agent under restricted configurations.
The goal is to determine which capabilities are essential for achieving robustness on C++ repositories and to quantify their individual contributions.

\textbf{RQ3: What are the behavioral characteristics of \toolname{} across its internal stages, specifically in terms of reproduction success and localization accuracy?}

This research question investigates the internal behavior of the system by examining two key metrics. 
The first is the reproduction success rate, which measures the ability of the Reproducer Agent to construct an executable test that triggers the issue. 
The second is the localization success rate, which quantifies whether the Patch Agent identifies the correct defect region before generating a patch. 
These measurements provide insight into where failures occur in the repair pipeline and help explain the end-to-end performance observed in RQ1.

\subsection{RQ1: Overall Issue Resolution Performance}

Table~\ref{tab:rq1_results} reports the end-to-end issue resolution rates of \toolname{}
and all baselines on \texttt{Multi\-SWE\-bench\-CPP}.  
\toolname{} equipped with GPT-5 achieves the highest overall performance, resolving
25.58\% of all issues, which represents a substantial improvement over prior systems.
The strongest baseline, MOpenHands + Claude-3.7 Sonnet, resolves 14.73\%, while
MSWE-agent and MAgentless achieve 11.63\% and 3.88\%, respectively.  
Thus, \toolname{} exceeds the strongest Claude-3.7-based baseline by 
10.85 absolute percentage points and achieves more than twice the resolution
rate of MSWE-agent.

When using DeepSeek-V3, \toolname{} obtains 13.20\% resolution accuracy,
again surpassing all DeepSeek-based baselines by a clear margin.  
MOpenHands, MSWE-agent, and M a gen t le s s with DeepSeek-V3 achieve 7.75\%, 7.75\%,
and 1.55\%, respectively.  
This demonstrates that \toolname{} consistently improves C++ repair performance
across different LLM backends.

A difficulty-level analysis further highlights the robustness of \toolname{}.
With GPT-5, \toolname{} resolves 50.00\% of easy issues, 25.42\% of medium issues,
and 9.52\% of hard issues, all of which exceed the best-performing baselines by substantial margins.  

\begin{table*}[t]
\centering
\small
\caption{Issue resolution rates on \texttt{Multi\-SWE\-bench\-CPP}. 
The InfCode models are compared with leading baselines using Claude 3.7 Sonnet and DeepSeek-V3. 
For settings where difficulty-level breakdown is not provided, we report ``--''.}
\label{tab:rq1_results}
\begin{tabular}{lcccc}
\toprule
\textbf{Model} & \textbf{Overall (\%)} & \textbf{Easy (\%)} & \textbf{Medium (\%)} & \textbf{Hard (\%)} \\
\midrule
\textbf{InfCode + GPT-5} & \textbf{25.58} & \textbf{50.00} & \textbf{25.42} & \textbf{9.52} \\
\textbf{InfCode + DeepSeek-V3} & \textbf{13.20} & -- & -- & -- \\
\midrule
\multicolumn{5}{l}{\textit{Baselines using Claude 3.7 Sonnet}} \\
Mopenhands + Claude-3.7-Sonnet & 14.73 & 32.14 & 15.25 & 2.38 \\
MSWE-agent + Claude-3.7-Sonnet & 11.63 & 28.57 & 11.86 & 0.00 \\
MAgentless + Claude-3.7-Sonnet (Oct) & 3.88 & 10.71 & 3.39 & 0.00 \\
\midrule
\multicolumn{5}{l}{\textit{Baselines using DeepSeek-V3}} \\
Mopenhands + DeepSeek-V3 & 7.75 & 10.71 & 10.17 & 2.38 \\
MSWE-agent + DeepSeek-V3 & 7.75 & 17.86 & 8.47 & 0.00 \\
MAgentless + DeepSeek-V3 & 1.55 & 3.57 & 0.00 & 0.00 \\
\bottomrule
\end{tabular}
\end{table*}

\begin{figure*}[t]
\centering
\begin{minipage}{0.48\linewidth}
    \centering
    \includegraphics[width=\linewidth, trim=10pt 25pt 10pt 25pt, clip]{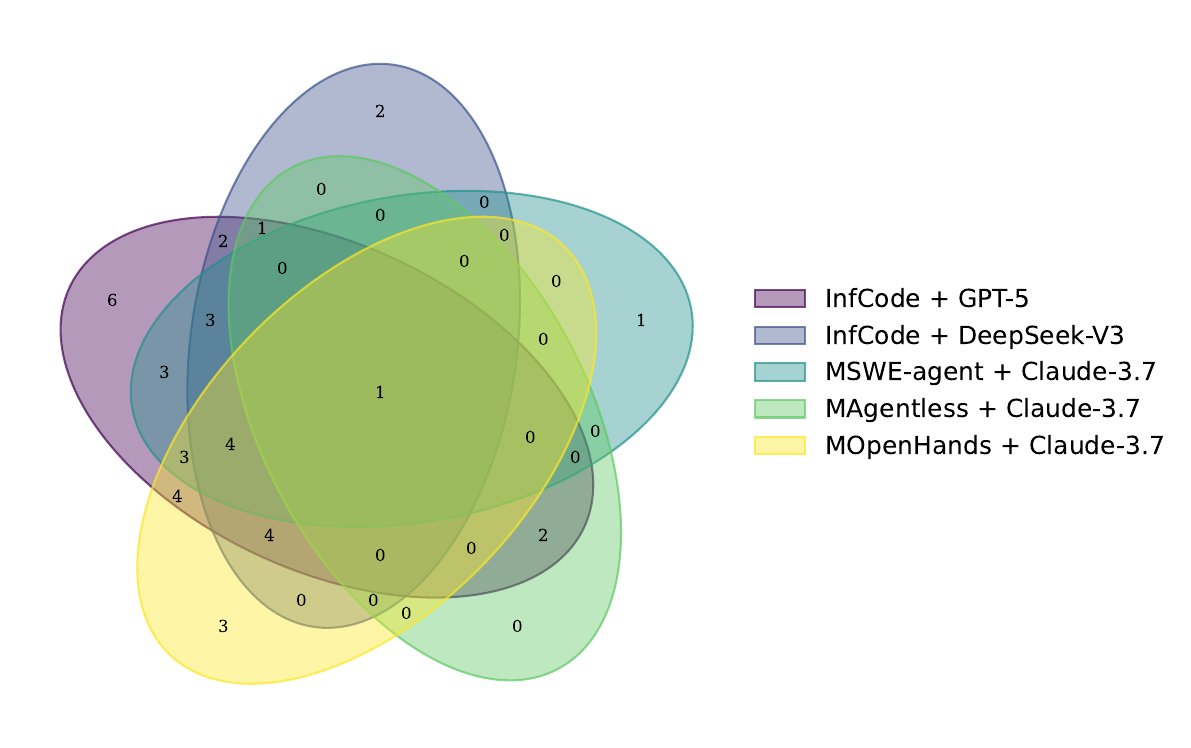}
    \caption{Instance-level overlap of resolved issues on \texttt{Multi\-SWE\-bench\-CPP} across InfCode and representative baselines.}
    \label{fig:venn5}
\end{minipage}
\hfill
\begin{minipage}{0.48\linewidth}
    \centering
    \small
    \captionof{table}{Ablation study of \toolname{} on \texttt{Multi\-SWE\-bench\-CPP}. 
    All configurations retain the Patch Agent; individual retrieval, reasoning, and validation components are removed to quantify their contributions.}
    \label{tab:ablation}
    \vspace{4pt}
    \begin{tabular}{l c}
    \toprule
    \textbf{Configuration} & \textbf{Resolution (\%)} \\
    \midrule
    \textbf{Full System (GPT-5)} & \textbf{25.58} \\
    \midrule
    \textit{Ablations (Patch Agent retained)} & \\
    \quad w/o Code-Intent Retrieval & 19.37\% \\
    \quad w/o AST-Structured Querying & 17.05\% \\
    \quad w/o Reproducer Agent & 20.16\% \\
    \quad w/o Selector Agent & 22.48\% \\
    \bottomrule
    \end{tabular}
\end{minipage}
\end{figure*}

For example, on hard problems, the most structurally complex C++ issues involving
deep template instantiation chains or multi-file dependency interactions—MOpenHands + Claude-3.7 achieves only 2.38\%, whereas \toolname{} achieves 9.52\%.

Beyond aggregate resolution rates, we further analyze the instance-level repair
coverage of \toolname{} compared with representative baselines.
Figure~\ref{fig:venn5} presents a five-way Venn diagram spanning InfCode
(GPT-5 and DeepSeek-V3) and three Claude-3.7-based agents (MOpenHands,
MSWE-agent, and MAgentless).

InfCode + GPT-5 not only subsumes the majority of issues repaired by all
baselines but also contributes a non-trivial set of uniquely solved instances, 
6 issues are repaired exclusively by InfCode + GPT-5, and none of the other
four systems are able to fix them.
This demonstrates that \toolname{} expands the solvable region of the benchmark
rather than merely overlapping with existing capabilities.

Overall, the results demonstrate that \toolname{} provides significant and consistent improvements over existing agent-based repair systems. This confirms that the proposed
semantic and structural retrieval mechanisms are effective for resolving real-world C++ issues, which pose substantially greater localization and reasoning challenges than their Python counterparts.

\subsection{RQ2: Contribution of Retrieval and Reasoning Components}

Table~\ref{tab:ablation} presents the ablation study for \toolname{} on
\texttt{Multi\-SWE\-bench\-CPP}.  
All variants retain the Patch Agent so that the differences in performance reflect the contribution of the retrieval, reproduction, and validation
mechanisms rather than the underlying patch synthesis capability.


Removing semantic code-intent retrieval reduces the resolution rate from
25.58\% to 19.37\%, an absolute drop of 6.21 percentage points.  
This demonstrates that high-level semantic filtering is crucial for constraining
the search space in large C++ repositories. Since many issues lack direct fault
indicators, the absence of semantic retrieval deprives the Patch Agent of coherent
feature-level context, significantly weakening its ability to identify the
relevant defect region.

Disabling AST-structured querying yields an even larger absolute reduction,
with performance dropping from 25.58\% to 17.05\%, a decrease of 8.53 points.  
This confirms that structural code search is a core requirement for C++ repair.  
C++ projects contain deeply nested class hierarchies, overloaded member functions,
template instantiations, all of which introduce
structural ambiguity that lexical search cannot resolve.  
AST-based querying provides deterministic disambiguation of these constructs,
enabling precise localization and yielding a larger performance impact than
semantic retrieval alone.

Removing the Reproducer Agent decreases performance to 20.16\%,
a 5.42 points drop.  
Without a faithful executable test derived from the issue description, the system
loses the behavioral specification that constrains the localization process and
validates patch candidates. This leads to higher rates of both under-constrained
patch generation and false positives that would otherwise be rejected.

Finally, removing the Selector Agent reduces resolution accuracy to 22.48\%,
an absolute loss of 3.10 points.  
The Selector Agent performs structured patch pruning by removing redundant edits,
filtering semantically equivalent or comment-only modifications, and applying
LLM-based ranking to choose the most behaviorally consistent candidate.  
The degradation shows that these verification and ranking steps meaningfully
increase the likelihood that the final patch is correct.

Overall, the ablation results indicate that the performance improvements reported
in RQ1 arise from the combined effect of semantic retrieval, structural analysis,
reproducer construction, and selector-guided filtering.  
Among these, semantic intent retrieval and AST-structured querying contribute the
largest absolute gains, demonstrating their necessity for robust repair in
complex C++ codebases.

Beyond correctness, we also examine the effect of removing retrieval
components on the reasoning efficiency of the Patch Agent.  
We measure efficiency using the number of LLM iterations required to
synthesize a patch.  
In the full system, the Patch Agent requires an average of 28.1 turns.
Removing semantic code-intent retrieval increases the average iteration
count to 35.3 turns, while removing AST-structured querying leads to a
substantial increase to 45.3 turns.

These results indicate that both retrieval mechanisms not only improve
repair accuracy but also significantly reduce the reasoning burden on
the LLM.  
Without semantic retrieval, the agent must explore a substantially
larger portion of the repository before reaching a plausible localization
hypothesis.  
Without AST-based structural analysis, the agent increasingly relies on
trial-and-error reasoning to resolve ambiguities in overloaded members,
template instantiations, and cross-file dependencies.  
Thus, retrieval-guided reasoning is essential not only for accuracy but
also for maintaining tractable search trajectories in C++ repair tasks.

\subsection{RQ3: Behavioral Analysis of Internal Pipeline Stages}

To understand where failures occur within the repair pipeline, we analyze the
internal behavior of \toolname{} across three stages:  
(1) reproduction of the issue,  
(2) localization of the defect, and  
(3) final end-to-end resolution.  
Table~\ref{tab:rq3_behavior} summarizes the results.

\paragraph{Reproduction Success Rate.}
The Reproducer Agent successfully constructs an executable failing test for
28.81\% of issues.  
This indicates that more than half of the C++ issues cannot be reproduced from
the textual description alone.  
Compared to Python-based benchmarks, C++ issues often rely on build-system
configurations, platform-specific behaviors, or cross-module runtime
states that are not explicitly described in the issue report.  
Consequently, missing environment assumptions and implicit dependencies make
reproduction failures a major limiting factor in C++ repair.

\paragraph{Localization Success Rate.}
Localization performance is analyzed at two granularities.  
At the file level, the Patch Agent identifies the correct file containing the
defect for 55.10\% of issues.  
At the function level, accuracy decreases to 42.10\%, reflecting the increased
difficulty of pinpointing defects in fine-grained C++ program structures.  
This gap highlights the inherent ambiguity in C++ codebases, where overloaded
member functions, template instantiations, namespace shadowing, and deep
inheritance hierarchies obscure semantic relationships that are more explicit in
languages like Python.  
Even with semantic code-intent retrieval and AST-structured querying,
mislocalization remains the most common failure source after reproduction.

\paragraph{End-to-End Success.}
Ultimately, \toolname{} resolves 25.58\% of issues, as reported in RQ1.
By correlating the three behavioral metrics, we observe a clear pattern:
(1) issues that fail to reproduce cannot proceed further in the pipeline, and  
(2) issues that reproduce but mislocalize seldom yield correct patches, 
even when patch synthesis succeeds locally.  
Thus, the primary failure modes lie in reconstructing the execution context and 
identifying the correct structural region of the defect, rather than in the
patch generation step.

Overall, the behavioral analysis demonstrates that C++ repair presents
fundamental challenges that differ significantly from Python-based settings.
Reproduction requires inferring complex runtime contexts, while localization
requires resolving structural ambiguities intrinsic to the C++ language.
The results confirm that \toolname{}’s retrieval and structural reasoning
components directly target the most difficult stages of the pipeline, enabling
substantial improvements in end-to-end repair capability.

\begin{table}[t]
\centering
\small
\caption{
Behavioral analysis of \toolname{} on \texttt{Multi\-SWE\-bench\-CPP}. 
We report the success rate of each internal stage in the repair pipeline.
}
\label{tab:rq3_behavior}
\begin{tabular}{l c}
\toprule
\textbf{Metric} & \textbf{Rate (\%)} \\
\midrule
Reproduction Success Rate 
& 28.81 \\
File-Level Localization Success Rate 
& 55.10 \\
Function-Level Localization Success Rate 
& 42.10 \\
End-to-End Resolution Rate 
& 25.58 \\
\bottomrule
\end{tabular}
\end{table}

\section{RELATED WORK}

\subsection{Automatic Software Issue Resolution}

Automatic software issue resolution has received significant attention in recent years, leading to a growing body of work centered around LLM-driven agents~\cite{openhands,agentless,dars,sweagent,autocoderover,bugpilot,experepair,swedebate,sweexp,lingmaagent,trae,specrover,repograph,test-time-scaling,swegpt}. Early agent frameworks such as openhands~\cite{openhands} introduced a tool-augmented planning mechanism that enables iterative interaction with the environment through file editors, terminals, and search tools. SWE-agent~\cite{sweagent} extends this idea by providing a dedicated agent--computer interface tailored for repository-level modification, allowing the agent to inspect, edit, and validate code through structured operations.

Subsequent systems focus on improving retrieval and localization. Moatless~\cite{moatless} enhances issue resolution by incorporating code search utilities and retrieval strategies designed to identify potentially relevant locations within a Python codebase. AutoCodeRover~\cite{autocoderover} introduces a structural representation of source code based on abstract syntax trees (ASTs), enabling more precise localization compared with lexical search. Building on this design, SpecRover~\cite{specrover} integrates function-level summarization and automatic reproduction test generation to improve localization and verification fidelity. Agentless~\cite{agentless} adopts a fixed operational pipeline—fault localization, patch generation, and patch validation—without requiring tool-based exploration, while MarsCode Agent~\cite{liu2024marscode} employs code knowledge graphs combined with LLM reasoning to support localization and candidate patch generation.

Beyond retrieval-oriented approaches, several recent systems explore alternative 
mechanisms to enhance agent robustness and patch quality. 
SWE-EXP~\cite{sweexp} and EXPERepair~\cite{experepair} incorporate long-horizon historical memory, enabling agents 
to accumulate and leverage prior interaction trajectories to guide future repair attempts. 
SWE-Debate~\cite{swedebate} introduces a multi-agent debate protocol in which multiple LLMs generate 
competing reasoning chains and critique each other, improving the likelihood that the system converges on the correct repair trajectory. 
Trae~\cite{trae} further advances this direction by combining test-time scaling~\cite{test-time-scaling} with 
a selector agent that evaluates candidate patches using behavioral signals, achieving state-of-the-art results on the SWE-bench-verified~\cite{openai_swe_bench_verified_2024}  benchmark. 
These techniques highlight the growing shift toward iterative, ensemble-based, or 
interaction-driven reasoning; however, they remain optimized for Python environments 
and do not address the structural and semantic complexities inherent to C++ codebases.

Despite architectural differences, all of these systems share a fundamental characteristic: they have been designed, evaluated, and optimized almost exclusively for \emph{Python} repositories. The benchmarks used to validate them, most notably SWE-bench~\cite{swebench}, are dominated by dynamically typed, memory-safe Python projects. Under these conditions, state-of-the-art systems achieve strong performance; for example, MOpenHands combined with Claude 3.7 Sonnet~\cite{anthropic_claude_sonnet3.7_2025} resolves 52.2\% of SWE-bench Python issues. However, when evaluated on the C++ subset of MultiSWE-bench, the same configuration achieves only 14.73\%. This sharp degradation demonstrates that general-purpose agent designs do not transfer effectively to statically typed, structurally complex languages.

The difficulty arises from the nature of C++ codebases. Identifier overloading, namespace shadowing, template instantiation, deep inheritance hierarchies, and cross-module control flow cause simple lexical retrieval---even when paired with powerful LLMs---to produce ambiguous or misleading context. Python-oriented agents typically rely on heuristics such as keyword matching, directory-level search, or shallow name-based filtering. These retrieval mechanisms are inadequate for C++, where the same identifier may denote distinct functions, templates, or methods across multiple nested namespaces, and where the syntactic and semantic relationships required for fault localization cannot be recovered from surface-level cues.

Furthermore, the majority of prior work assumes the availability of explicit execution traces or easily inferable failure points. In C++, non-crashing logic bugs frequently lack observable fault indicators; without symbolic traces or stack information, agents must infer the fault location solely from the semantic structure of the repository. Existing Python-oriented systems are not equipped to perform this form of structural reasoning.

\subsection{Ensemble Methods for LLM-based Systems}

Ensemble learning techniques~\cite{polikar2012ensemble,sagi2018ensemble} have been widely used to improve robustness and generalization by aggregating diverse predictions. In the context of LLM-based systems, ensemble strategies have recently demonstrated strong performance gains. Studies in mathematical reasoning, such as Best-of-N~\cite{snell2025scaling}, show that selecting among multiple LLM-generated solutions via a scoring model yields substantial improvements. In competitive programming, S*~\cite{li2025s} integrates execution feedback from public and LLM-generated tests with clustering-based selection to identify high-quality code candidates. EnsLLM~\cite{mahmud2025enhancing} combines syntactic similarity (CodeBLEU~\cite{ren2020codebleu}) and behavioral similarity (CrossHair~\cite{schanely2025crosshair}) to vote among candidate patches.

Several works adapt ensemble-style selection to software issue resolution. Augment~\cite{augment2025swebench} adopts an LLM-as-a-judge strategy to evaluate semantic alignment between issue descriptions and candidate patches. DeiBase~\cite{zhang2024diversity} prompts the LLM to generate explanations and confidence scores for each patch before ranking candidates. While effective in Python benchmarks, these ensemble techniques rely on the underlying agent’s ability to retrieve accurate code context---a capability that degrades severely in C++ environments.

\subsection{Summary and Positioning}

Existing approaches demonstrate substantial progress on Python issue resolution but do not address the structural complexity inherent in C++. As evidenced by the substantial performance decline observed when Python-optimized agents are applied to C++ repositories, the dominant research direction lacks mechanisms for resolving identifier ambiguity, disambiguating overloaded declarations, or navigating deep syntactic structures. Consequently, these systems are unable to perform reliable localization or context retrieval in large-scale C++ codebases.

In contrast, \toolname{} is designed specifically to address these deficiencies. By integrating semantic code-intent retrieval with deterministic AST-based querying, \toolname{} provides a C++-aware retrieval and localization pipeline that overcomes the limitations of Python-focused agents. This enables reliable context construction and significantly improves issue resolution in complex C++ repositories.

\section{THREATS TO VALIDITY}

Several factors may affect the validity of our findings. 
Internal validity may be influenced by the nondeterminism of LLM generation, though we mitigate this by using fixed seeds where possible and executing all agents under identical compute budgets. 
Construct validity is limited by the evaluation protocol of MultiSWE-bench~\cite{multiswebench}, where issue resolution rate does not capture partial progress, and a minority of issue reports include stack traces that may slightly simplify reproduction. 
External validity is constrained by the use of C++ projects from a single benchmark; results may differ on other domains such as embedded systems or mixed-language repositories. 
Nevertheless, the consistent improvements observed across all settings suggest that the benefits of \toolname{}’s C++-aware retrieval and structured reasoning generalize beyond the evaluated projects.

\section{CONCLUSION}

We presented \toolname{}, a C++-aware autonomous system for repository-level issue resolution. 
Unlike prior agents designed for Python, \toolname{} integrates semantic code-intent retrieval and AST-structured querying to address the identifier ambiguity and deep structural complexity of C++ codebases. 
Evaluated on \texttt{MultiSWE-bench-CPP}, \toolname{} achieves state-of-the-art performance, resolving 25.58\% of issues and substantially outperforming leading baselines. 
Ablation and behavioral analyses highlight the importance of both semantic and structural retrieval, and reveal current bottlenecks in reproduction and localization. 
Overall, \toolname{} demonstrates that reliable C++ repair requires language-aware reasoning beyond existing agent designs and provides a foundation for future multi-language issue resolution systems.

\bibliographystyle{IEEEtran}
\bibliography{reference.bib}

\end{document}